\documentclass[]{bmcart}

\usepackage[utf8]{inputenc} 

\usepackage{graphicx}

\begin{document}

\begin{frontmatter}

\begin{fmbox}


\title{Spreading of performance fluctuations on real-world project networks}


\author[
   addressref={aff1,aff2},                   
   corref={aff1},                       
   email={iacopo@nodeslink.com}   
]{\inits{IP}\fnm{Iacopo} \snm{Pozzana}}
\author[
   addressref={aff1},
   email={christos@nodeslink.com}
]{\inits{CE}\fnm{Christos} \snm{Ellinas}}
\author[
   addressref={aff1},
]{\inits{GK}\fnm{Georgios} \snm{Kalogridis}}
\author[
   addressref={aff1},
]{\inits{KS}\fnm{Konstantinos} \snm{Sakellariou}}


\address[id=aff1]{
  \orgname{Nodes \& Links Ltd}, 
  \street{107 Fleet Street},                     %
  \postcode{EC4A 2AB}                                
  \city{London},                              
  \cny{UK}                                    
}
\address[id=aff2]{%
  \orgname{Birkbeck, University of London},
  \street{Malet Street},
  \postcode{WC1E 7HX}
  \city{London},
  \cny{UK}
}


\begin{artnotes}
\end{artnotes}

\end{fmbox}


\begin{abstractbox}

\begin{abstract} 
Understanding the role of individual nodes is a key challenge in the study of spreading processes on networks. In this work we propose a novel metric, the reachability-heterogeneity (RH), to quantify the vulnerability of each node with respect to a spreading process on a network. We then introduce a dataset consisting of four large engineering projects described by their activity networks, including records of the performance of each activity; such data, describing the spreading of performance fluctuations across activities, can be used as a reliable ground truth for the study of spreading phenomena on networks. We test the validity of the RH metric on these project networks, and discover that nodes scoring low in RH tend to consistently perform better. We also compare RH and seven other node metrics, showing that the former is highly interdependent with activity performance. Given the context agnostic nature of RH, our results, based on real-world data, signify the role that network structure plays with respect to overall project performance.

\end{abstract}




\end{abstractbox}
%

\end{frontmatter}



\section*{Introduction}
Spreading broadly refers to the notion of an entity propagating through a networked system, typically fueled by a dynamical process \cite{pastor2015epidemic}.
Spreading processes are a powerful set of tools for modelling a wide-range of real-world phenomena, including the dissemination of (dis)information on social media \cite{vosoughi2018spread}, the propagation of a pathogen within a population \cite{santolini2018predicting}, cyber attacks on computer networks \cite{cohen2003efficient} and delays in transportation systems \cite{preciado2014optimal}.
Node degree \cite{wasserman1994social}, betweenness centrality \cite{freeman1977set} and eigenvector centrality \cite{bonacich1972factoring} are all examples of topological metrics used to approximate the role of individual nodes in the context of spreading processes, a problem that yet remains open in the extant literature \cite{radicchi2016leveraging, erkol2018influence}.

The problem is further complicated by the scarcity of reliable ground truth. Datasets providing an individual-level description of a spreading process within a population are few \cite{groendyke2011bayesian, chinazzi2020effect}, with aggregated reports being more common \cite{stack2013inferring}.
Even when working with real-world networks, researches often resort to simulations for what concerns the spreading dynamics itself \cite{mishra2016impact, davis2020phase};
 and when information describing the network structure is also incomplete, the interplay between the two problems further amplifies the difficulty of the task \cite{gomez2012inferring}.
 
 A bountiful, yet underexploited, source of reliable data, describing both complete network structures and the fine-grained evolution of real spreading processes on them, can be found within the field of project management \cite{ellinas2016project, vanhoucke2013overview, santolini2020uncovering}.
 Projects are described by schedules, time-ordered lists of interconnected activities that can be naturally modelled as directed acyclic graphs (DAGs) \cite{valls2001criticality}.
 
 Spreading can be used to describe performance fluctuations on project networks: activities completed behind or ahead of schedule can impact other activities downstream and initiate a spreading process \cite{ellinas2015robust, guo2019modeling}.
 Project schedules record both planned and real starting dates for all activities, therefore providing a complete record of the performance fluctuation dynamic.
 
 Real-world projects often perform poorly in terms of both time and cost, a fact that holds true across different countries, companies, and industries \cite{evrard2004boosting, budzier2011your}.
 As an example, studies have shown that, in the construction sector, almost nine out of ten projects are subject to cost overruns, for an average overrun cost estimated to be as high as 45\% \cite{flyvbjerg2003common, flyvbjerg2007cost}.
 
 Large failures in projects often start as localised phenomena, with the performance of a single activity eventually impacting the performance of the entire project.
 Cases have been documented where an initial disruption located in a single activity ended up affecting almost a third of the entire project \cite{sosa2014realizing}, or increasing its final cost by 20 to 40\% \cite{terwiesch1999managing}.
 In this respect, the networked structure of the schedule has been shown to play an important role \cite{ellinas2019domino, mihm2003problem}.
 
 Methodologically, most of the efforts aimed at modelling project performance through their associated networks have centered on cascade models \cite{wang2018development}, for example by focusing on how small-scale delays can trigger project-wide cascades \cite{ellinas2019domino, santolini2020uncovering}, or by studying the role of indirect interactions between activities \cite{ellinas2018modelling}.
 With the present study, we contribute to this line of work by developing a measure that draws a direct connection between topology and performance at the activity level, and then validate it using real performance data.
 
 Our contribution is two-fold. First, building on prior work by Estrada \cite{estrada2010quantifying} and by Ye and colleagues \cite{ye2013entropy}, we introduce a novel measure called reachability-heterogeneity (RH), which quantifies heterogeneity on DAGs.
 The RH is defined both at the global (how heterogeneous is a network) and local level (how much a node contributes to the heterogeneity).
 
 Heterogeneity plays an important role in determining how vulnerable a network is with respect to spreading processes \cite{moreno2002epidemic}.
 If all nodes have equal spreading power, then the network is maximally robust, not presenting any weak spots to either targeted attacks or random failures \cite{xiao2018correlation}.
 Numerous studies quantify heterogeneity by examining the distribution of some node-level measure (examples including degree \cite{sun2016impact}, memory \cite{karsai2014time, sun2015contrasting}, activity potential \cite{perra2012activity, liu2014controlling}, attractiveness \cite{pozzana2017epidemic}, burstiness \cite{ubaldi2017burstiness} and modularity \cite{nadini2018epidemic}), and examine the relationship between such heterogeneity and the spreading dynamics. 

The novelty of our contribution consists in leveraging a topological feature that is intrinsically related to the spreading process: the number of descendants and of ancestors. Due to the absence of cycles, the size of the ancestry trees plays an especially important role in DAGs; and, to the best of our knowledge, there is no study examining the relevance of its heterogeneity in spreading processes. Our analysis qualitatively verifies that the global RH score is a good indicator of the heterogeneity of the ancestry and descendancy distributions.

Our second contribution consists in the introduction of a dataset describing the networks of activities that make up four real-world, complex projects; these data provide a reliable ground truth for benchmarking spreading processes. We experimentally validate the accuracy of RH against performance records from the projects’ activities. Our results show that best-performing nodes tend to score low in RH, making our metric a good tool for their identification. Furthermore, we compare the local RH to seven other node metrics by computing the mutual information between them and the activity performance; RH reports the highest (or, in one case, third-highest) mutual information values among all candidates. Given the context agnostic nature of RH, our results signify the role that the network structure has with respect to overall project performance, and indicate that the RH score gives computational embodiment to the notion that a network is maximally robust against spreading when all nodes contribute equally to it.

\section*{Data and Methods}
\subsection*{Project Data}

We use data from four complex engineering projects, where `complex’ refers to the non-triviality of underlying dependencies \cite{baccarini1996concept, jacobs2011product, ellinas2016toward}.
For each project, we use the schedule to generate the respective activity network \cite{valls2001criticality}.
The project schedule consists of a list of activities and in a list of dependencies between them. For each activity, the schedule contains the planned and actual start and end date.
For each activity, the schedule contains the planned and actual start and end date.
Target dates for an activity correspond to its start and end date as initially planned.
Actual dates, as the name suggests, correspond to the dates when the activity was actually initiated and completed.

The schedule naturally lends itself to be represented as a network, with activities taking the role of nodes and dependencies representing directed links among them (from now on, we will use the terms `node’ and `activity’ interchangeably).
A link from node $i$ to node $j$ means that activity $i$ must first be completed before activity $j$ can start.
At this stage, we remove from the network all isolated nodes, since these nodes are not capable of contributing to any sort of spreading in a meaningful way.
Notice that activity networks are DAGs, as cyclic dependencies between activities are not allowed.

The four projects analysed here detail the construction of different kinds of infrastructure: a  highway (HW), a data centre (DC), a wind farm (WF) and a power network (PN).
The number of activities and dependencies for each project ranges from less than two hundred to more than a thousand (Table \ref{data_tab}).
Activity networks do not necessarily consist of a single component: projects may have a modular structure, being composed of independent sections.
The number of weakly connected components for each network, and the size of the largest one, are also reported in Table \ref{data_tab}.
We verify that all four networks are acyclic, as expected.

\begin{table}
\caption{For each of the four activity networks we report the number of activities (nodes), dependencies (directed links) and weakly connected components, and the size of the largest weakly connected component.}
      \begin{tabular}{lcccc}
        \hline
        Project & Activities & Dependencies & WCCs & LWCC\\ \hline
        Highway (HW) & 682 & 666 & 113 & 100 \\
        Data Centre (DC) & 1185 & 1510  & 111 & 440 \\
        Wind Farm (WF) & 266  & 425 & 1 & 266 \\
        Power Network (PN) & 129  & 138 & 10 & 62 \\ \hline
      \end{tabular}
      \label{data_tab}
\end{table}

\begin{figure}
  \includegraphics[width=0.9\textwidth]{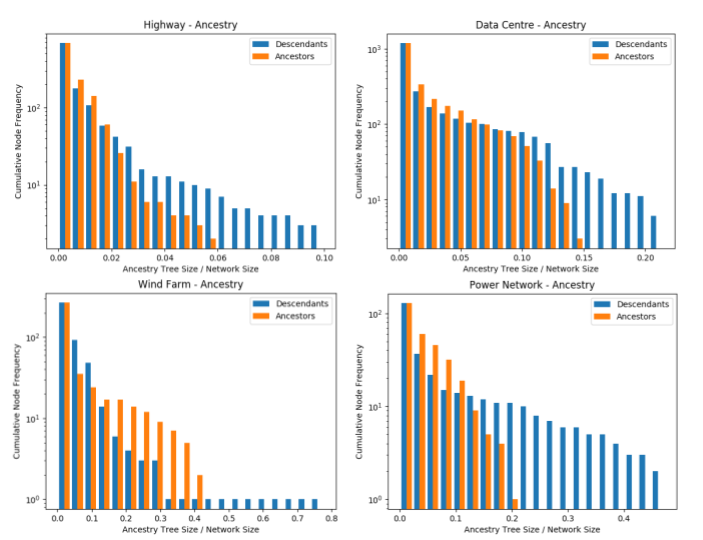}
  \caption{Reverse cumulative frequency distribution of the fraction of descendants (blue) and ancestors (orange) over the total number of nodes. The distributions vary widely in terms of largest ancestry and descendancy fraction (from less than $0.1$ for HW, to more than $0.7$ for WF), showing different degrees of heterogeneity.}
      \label{desc_fig}
      \end{figure}

Figure \ref{desc_fig} shows the reverse cumulative distribution of the number of ancestors and descendants for each project network, divided by the network’s size.
The four dataset present significant differences between each other, with the most peaked (HW) having no ancestry or descendancy larger than $0.1$, while WF and PN have numerous nodes with either descendancy or ancestry ranging between $0.2$ and $0.5$ of the entire network.
In all cases the distribution of descendants has the longest tail of the two, although in the case of WF this is caused by the presence of a single node with a large number of descendants (more than $0.7$ of all nodes).
Overall, the four datasets show very different degrees of heterogeneity in their ancestry and descendancy distributions.

\subsection*{Activity Performance}

Performance indicators for each activity can be constructed by comparing its target with the actual start and end dates.
Here we focus on a particular form of performance, the Start Delay \emph{i.e.}, the difference between the target and the actual start date.
The advantage of this metric is that it allows us to focus on performance fluctuations that occurred upstream of an activity, separating them from fluctuations that might occur while the activity is being carried out.
A possible alternative performance indicator would be represented by the End Delay, \emph{i.e.}, the delay in the end date of an activity; this second measure would account for fluctuations that occur while the activity is taking place too, as well as for those that took place upstream.

Suppose, for example, that the completion of activity $j$ is dependent on the completion of activity $i$, and the two activities are taking place at the same time.
If a delay happens in $i$ after the start of $j$, the same delay might end up propagating to $j$ as well, delaying its completion; therefore the End Delay would capture such propagation, while the Start Delay would not.
However, a significant downside of the End Delay is that it also accounts for the emergence of performance fluctuations within the activity itself (endogenous fluctuations), \emph{i.e.}, fluctuations that would have occurred even if the activity had taken place in isolation, and that are, hence, independent of the network topology; by using End Delay alone, it is impossible to disentangle the two types of phenomena. We therefore focus on the Start Delay as our performance metric. 

\begin{figure}
  \includegraphics[width=0.9\textwidth]{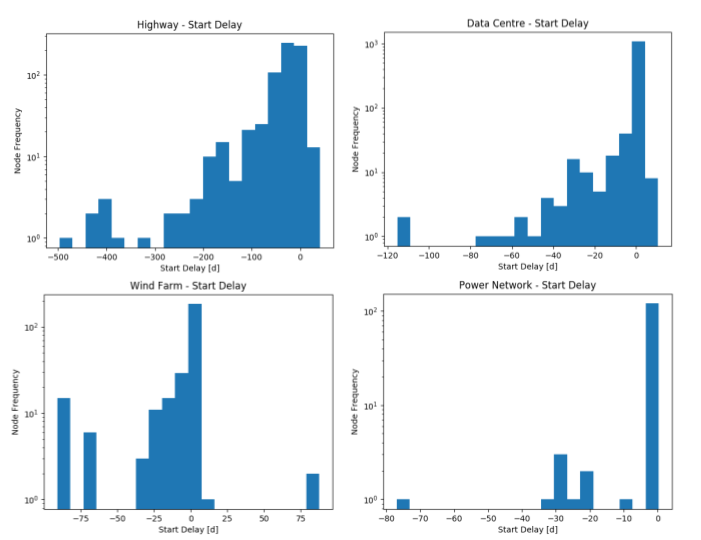}
  \caption{Frequency distribution of Start Delay (in days) for different activities. All distributions are starkly peaked around zero, with values close to the peak surpassing their further counterparts by orders of magnitude (hence the need for the logarithmic scale). HW and DC show a left tail, and WF is the only dataset recording delays larger than a few units.}
      \label{perf_fig}
      \end{figure}

In Figure \ref{perf_fig}, we plot the distribution of Start Delay values, measured in days.
Most recorded values are negative, indicating that an activity has started ahead of schedule.
Only in WF values larger than a few (positive) units appear.
In all cases, the distribution peaks at zero, corresponding to activities having started as planned.
HW and DC show a distinct left tail, with the frequency of activities decreasing as the Start Delay decreases.
In all four cases, frequencies range over several orders of magnitude, warranting the use of a logarithmic scale on the y-axis.
Notice that a majority of activities starting early (i.e., with negative delay) does not necessarily result in an early completion of the project as a whole.
As discussed in the Introduction, project overruns are often driven by localised disruptions in activities that can trigger fluctuations reaching the project’s end, and affecting the overall performance.

\subsection*{Reachability-Heterogeneity Measure}

To quantify the heterogeneity of a project network, we start from Estrada’s heterogeneity measure \cite{estrada2010quantifying}, and particularly its extension to directed graphs \cite{ye2013entropy}:

\begin{equation} \label{direct_eq}
\rho(G) = \frac{1}{|N| - 2 \sqrt{|N| - 1}} \sum_{(i,j) \in E} \left( \frac{1}{\sqrt{k_i^{out}}} - \frac{1}{\sqrt{k_j^{in}}} \right)^2
\end{equation}

Above, $k^{in}_i$ and $k^{out}_i$ represent the in- and out-degree of node $i$ respectively, $N$ is the set of all edges in the network $G$, and the summation is taken over the set of all $G$'s (directed) edges $E$.

Since activity networks are DAGs, a performance fluctuation in node $i$ can only propagate to its descendants.
In turn, node $i$ can only be affected by performance fluctuations occurring in its ancestors.
By descendant of $i$, we mean any node $j$ such that a directed path from $i$ to $j$ exists; by ancestor of $i$, we mean any node $j$ such that a directed path from $j$ to $i$ exists. $i$ is a descendant of $j$ if and only if $j$ is an ancestor of $i$.

In assessing the heterogeneity of an activity network with respect to performance fluctuation spreading, we make use of the more cogent notion of ancestor (descendant) instead of predecessor (successor).
The contribution of a pair to the overall score is a function of the difference between the number of ancestors and descendants of the two nodes involved, rather than of their in- and out-degree, accounting for the impact of ancestors and descendants to the overall spreading process.

In formulae, we replace the in- and out-degree from Equation \ref{direct_eq} with the number of ancestors and descendants of the two nodes respectively, and we extend the summation to all pair of connected nodes, leading to the following definition:

\begin{equation} \label{dag_eq}
RH^{global}(G) = \frac{1}{|N| - 2 \sqrt{|N| - 1}} \sum_{(i,j) \in C} \left( \frac{1}{\sqrt{d_i}} - \frac{1}{\sqrt{a_j}} \right)^2
\end{equation}

In Equation \ref{dag_eq}, $d_i$ and $a_i$ represent the number of descendants and ancestors of node $i$, and $C$ is the set of all ordered pairs of connected nodes.
This metric is a \emph{global} network property that allows comparison between different topologies and quantification of their heterogeneity with respect to the size of nodal ancestry lineages. In comparison, the measure in Equation \ref{direct_eq} focuses exclusively on the immediate neighbourhood of the node.

In order to provide more actionable information, we introduce an additional version of the measure above, defined at the level of single nodes, in order to allow targeted interventions by project experts. Our aim in doing so is to answer the question: if a single node could be removed in order to make the topology less vulnerable, which one would be the best choice? The answer can simply be computed by taking the difference between the network scores before and after the removal:

\begin{equation} \label{local_eq}
RH^{local}(i) = RH^{global}(G) - RH^{global}(G \backslash \{i\})
\end{equation}

We call this measure Reachability-Heterogeneity (RH).

\section*{Results}
We first calculate the RH score for all nodes on all the four projects, as well as the four global RH scores, which are reported in Table \ref{rh_tab}.
The global score provides a good characterisation of the shape of the ancestry and descendancy distributions shown in Figure \ref{desc_fig}, with the highest RH value (WF) being assigned to the distribution with the longest tail, and the other three following in order.

\begin{table}
\caption{Global RH scores for the four activity networks. The comparison with Figure \ref{desc_fig} shows a correspondence between higher score values and longer tail in the ancestry tree size distribution.}
      \begin{tabular}{lc}
        \hline
        Project & Global RH \\ \hline
        Highway (HW) & 0.238 \\
        Data Centre (DC) & 0.332 \\
        Wind Farm (WF) & 0.680 \\
        Power Network (PN) & 0.514 \\ \hline
      \end{tabular}
      \label{rh_tab}
\end{table}

\begin{figure}
  \includegraphics[width=0.9\textwidth]{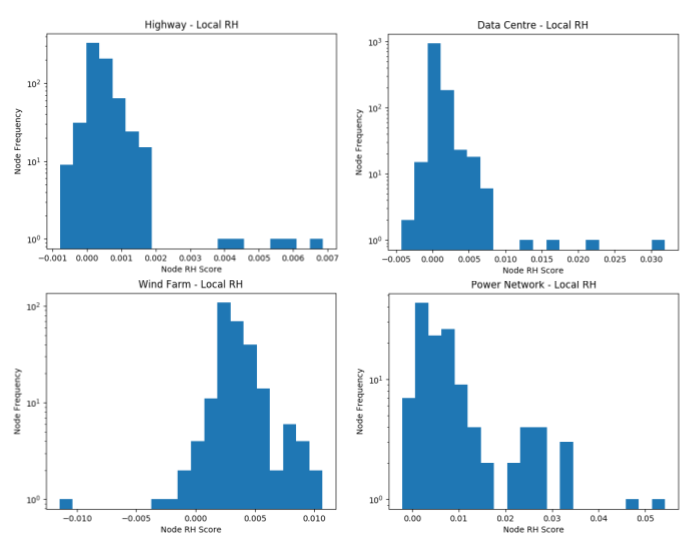}
  \caption{Distribution of local RH scores for the four activity networks. All four distributions have a clear peak, close to but not always coinciding with the zero value, with frequency values spanning over several orders of magnitude. WF is the only network exhibiting a left tail in the RH distribution, and comparison with Figure \ref{perf_fig} shows that it is also the only project that, among the four, reported delays significantly larger than zero.}
      \label{rh_fig}
      \end{figure}

The distributions of node-level RH scores for all four projects are shown in Figure \ref{rh_fig}. 
All distributions show frequency values spanning over various orders of magnitude and a rather clearly identifiable peak, always close, but not always corresponding, to the zero value. 
HW, DC and PN bear some degree of similarity in shape, with a single-sided flat tail in the higher values, but differ in magnitude. 
Interestingly, WF, which is the only project to report significant positive delays (Figure \ref{perf_fig}), is also the only project with a significant left tail in the RH score distribution; it is worth remarking that the RH score is based on the network structure alone, and does not account for performance data.

To assess the effectiveness of RH in quantifying node vulnerability, we first use activity performance to build our ground truth. 
Specifically, we use the Start Delay indicator, as described in the Methods section. To mitigate the noise, we group the nodes in bins of equal width.\footnote{We use the OptBinnig Python package to choose the number of bins:
http://gnpalencia.org/optbinning/.}
Within every bin, we calculate the Start Delay of each node and a number of summarising statistics, namely: mean, median, 50\% and 68\% Confidence Intervals (CIs). 

\begin{figure}
  \includegraphics[width=0.9\textwidth]{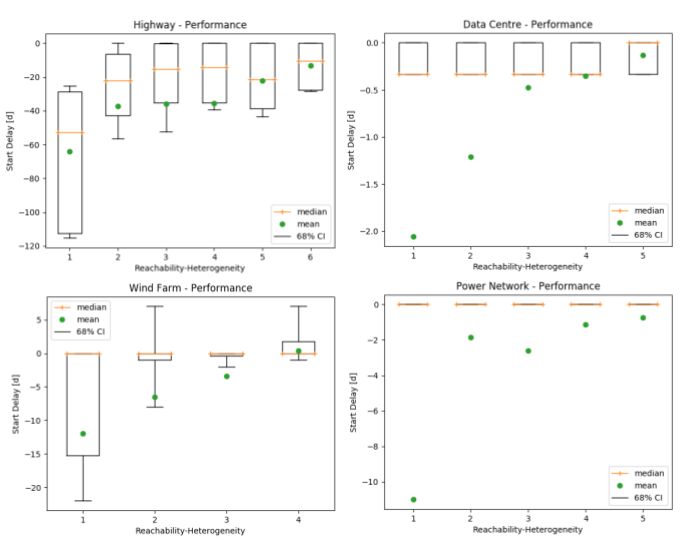}
  \caption{For each activity of each project, we report Start Delay (in days) and RH score (at the node level). Data are binned uniformly along the RH dimension to mitigate noise. A trend emerges in all four datasets with higher RH values corresponding to longer delays, \emph{i.e.}, worse performance. As it is particularly evident from DC and PN, a significant contribution to this phenomenon comes for the outliers in the Start Delay distribution, the best-performing activities, that tend to score low in RH.}
      \label{trend_fig}
      \end{figure}

The results for each project are reported in Figure \ref{trend_fig}, in the form of boxplots. In general, the Start Delay value increases for greater RH, showing that this newly introduced measure can provide a good indicator of activity performance. 
It is worth reminding that the Start Delay accounts for delays inherited from ancestors, signifying the relationship between performance and spreading (see the Data section for further discussion).

In particular, for the HW data the trend is especially evident in the mean and the lower end of the CIs. The upper end of the CIs seems to be capped at zero, as almost all Start Delay values are negative (see Figure \ref{perf_fig}). The trend is clearer for lower RH values,which then flattens towards the tail.

For the DC data, the trend is stronger in the mean. The clear separation between the mean value and the centre of the distribution confirms that Start Delay distributions within each bin are long-tailed, with longer tails in correspondence of lower RH values. Again, all Start Delay values are negative.

The WF data are the noisiest, possibly due to the smaller size of the dataset, leading to wider bins. Despite the noise, a trend, not captured by the median, can instead be seen in the CIs and mean.

Finally, in PN the same scenario as in DC is repeated, with the mean capturing a trend otherwise overlooked by the CIs, further reaffirming that low RH scores correspond to a greater presence of outliers from the (left) tail of the Start Delay distribution, the best-performing activities.
Due to the extremely peaked shape of the performance distribution (Figure \ref{perf_fig}), the small size of the CIs was indeed to be expected.

As a further step towards validating the effectiveness of the local RH score, we benchmark it against seven other node metrics: in-degree, out-degree, betweenness centrality, closeness centrality, reverse closeness (\emph{i.e.}, closeness centrality computed on the network with edges’ direction reversed), number of descendants and of ancestors.
Once again we use the Start Delay as our target variable.
For each of the eight metrics considered, we compute the mutual information between it and the target variable.\footnote{Notice that the notion of target variable has a purely methodological significance in this context: mutual information is symmetric with respect to the `candidate’ and `target’ distributions.}

For each of the four networks, we proceed by computing a two-dimensional frequency matrix with the considered node metric as one dimension and the Start Delay as the other.
For the purpose of computing frequencies, we group data in a number of uniform bins equal to the square root of the number of nodes, rounded down (the same number of bins is used along both dimensions). The mutual information is then computed through the frequency matrix.\footnote{More specifically, the mutual information is computed through the marginal and joint probability distributions for the two variables, as derived from the frequency matrix.}
The results, displayed in Table \ref{mi_tab}, show that the local RH has the highest mutual information value of all the metrics considered on all datasets minus DC, where it ranks third.

\begin{table}
\caption{Comparison between the local RH score and seven other node metrics. For every candidate, the table reports its mutual information score computed with the Start Delay as a target variable, and its rank in brackets. Local RH ranks third on DC and third on all other networks.}
      \begin{tabular}{lcccc}
        \hline
        Node Metric & Highway & Data Centre & Wind Farm & Power Network \\ \hline
In-degree & 0.287 (8) &
0.134 (7) & 0.285 (8) &
0.045 (8) \\
Out-degree & 0.304 (7) &
0.117 (8) & 0.293 (7) &
0.047 (7) \\
Betweenness & 0.920 (4) &
0.250 (6) & 0.667 (3) &
0.092 (6) \\
Closeness & 1.209 (2) &
\textbf{0.507 (1)} & 0.653 (4) &
0.106 (5) \\
Rev. Closeness & 0.975 (3) &
0.353 (4) & 0.689 (2) &
0.123 (4) \\
Descendants & 0.686 (6) &
0.274 (5) & 0.561 (6) &
0.148 (3) \\
Ancestors & 0.812 (5) &
0.382 (2) & 0.586 (5) &
0.149 (2) \\
Local RH & \textbf{1.709 (1)} &
0.354 (3) & \textbf{0.821 (1)} &
\textbf{0.208 (1)} \\
        \hline
      \end{tabular}
      \label{mi_tab}
\end{table}

\section*{Discussion}
Project performance can be understood by focusing on how fluctuations spread within the project’s underlying activity network.
We leverage the context agnostic nature of the approach to develop a new heterogeneity measure (RH) which we then use to explore four distinct projects (a highway, a data centre, a wind farm, and a power network respectively).
The size of the datasets varies between schedules too, from 1185 for DC to 129 for PN. 
The networks also have very different component structure, as summarised in Table \ref{data_tab}.

In all four cases, frequencies of ancestry size, descendancy size, and performance, take values ranging over various orders of magnitude. 
The global RH score (Table \ref{rh_tab}) appears to be particularly effective in quantifying the heterogeneity of the descendancy and ancestry distributions (Figure \ref{desc_fig}), with longer-tailed distributions (\emph{i.e}., more heterogeneous) corresponding to higher RH values. 
A systematic investigation of the nature of this correspondence is beyond the scope of this paper, and might provide the object of future works.

Our experimental results on the four datasets show that a general trend exists, according to which lower RH scores correspond to better performance (Figure \ref{trend_fig}). 
Looking at these results in detail, the cases of DC and PN are particularly interesting, with the mean of the binned data showing a clear trend that the median fails to capture. 
A similar behaviour is apparent in the other datasets too, though not as pronounced. This is due due to the trend being driven by outliers, \emph{i.e.}, best-performing activities, located in the left tail of the Start Delay distribution; these are activities that take smaller RH values and hence amplify the difference between mean and median values within each bin. 
Such a feature might prove convenient, considering that a likely purpose of the RH measure is to identify cases of extremely high performance, although the opposite (identifying the poorly performing nodes) might also be the case in some instances.

The use of the Start Delay as a performance measure allows us to draw a direct connection between performance and vulnerability to spreading, since it accounts for delays inherited from upstream nodes (as discussed in the Data section).
Three out of four projects (excluding WF) follow a similar Start Delay distribution, with a peak around zero and a tail in the negative values (corresponding to better-performing nodes).

As reported in Table \ref{mi_tab}, we run a comparison between the local RH score and seven other node metrics (in- and out-degree, betweenness centrality, closeness and reverse closeness centrality, number of descendants and of ancestors).
The purpose of the comparison is to quantify which of the candidate metrics carry the most information on node performance. 
To avoid making any assumption on the form of the dependency, we use mutual information, which is a non-parametric measure, capable of accounting for non-linear relationships. 
With the sole exception of DC, where it ranks third, the local RH carries the highest mutual information of all the metrics considered. 
No other candidate shows the same consistency across datasets; closeness centrality, for example, ranks first in DC and second in HW, but fourth in WF and fifth in PN. 
In- and out-degree are always the two worst performing metrics, reinforcing the point that an effective performance measure must look beyond the first-degree neighbourhood, in agreement with the existing literature \cite{lawyer2015understanding}.

\section*{Conclusions}
In the present work, we tackle the question of quantifying and mitigating spreading phenomena from a topological perspective, focusing on how fluctuations in the completion time of certain activities can impact the performance of complex projects. 
Our contribution is two-fold: first, we introduce a novel vulnerability measure that focuses on ancestry tree size, a quantity that plays a big role in spreading process across DAGs; second we apply this measure to an important but currently underrepresented domain - the delivery of complex projects - where we use ground truth data to test our proposed measure.

Using these data, we assess the effectiveness of RH in quantifying performance fluctuations of activities within projects. 
We show that higher values in RH correspond to worse performance, indicating its appropriateness in accounting for the propensity of such fluctuations to propagate. 
In addition, we compare RH with seven other node metrics, and show that RH carries the most amount of information about the activity performance on three out of four projects, strengthening its utility in identifying vulnerable nodes.

As well as introducing a new tool for the study of spreading processes on networks, and on directed acyclic graphs in particular, we hope that our work will stimulate the interest of the community in project management as a domain of application for network science.


\begin{backmatter}



\section*{Availability of data and materials}
The datasets analysed during the current study are not publicly available, in accordance with the terms under which access was granted by their owners to the authors, but are available from the corresponding author on reasonable request.

\section*{Competing interests}
  The authors declare that they have no competing interests.

\section*{Author's contributions}
IP, CE, KS and GK conceptualised the study, devised the methodology and collected the data. IP analysed the data. IP, CE, KS and GK wrote the manuscript. All authors read and approved the final manuscript.

\section*{Acknowledgements}
The authors would like to thank Stelios Avramidis for his valuable feedback regarding the manuscript.


\bibliographystyle{bmc-mathphys} 
\bibliography{bmc_article}      


\begin{thebibliography}{48}
\ifx \bisbn   \undefined \def \bisbn  #1{ISBN #1}\fi
\ifx \binits  \undefined \def \binits#1{#1}\fi
\ifx \bauthor  \undefined \def \bauthor#1{#1}\fi
\ifx \batitle  \undefined \def \batitle#1{#1}\fi
\ifx \bjtitle  \undefined \def \bjtitle#1{#1}\fi
\ifx \bvolume  \undefined \def \bvolume#1{\textbf{#1}}\fi
\ifx \byear  \undefined \def \byear#1{#1}\fi
\ifx \bissue  \undefined \def \bissue#1{#1}\fi
\ifx \bfpage  \undefined \def \bfpage#1{#1}\fi
\ifx \blpage  \undefined \def \blpage #1{#1}\fi
\ifx \burl  \undefined \def \burl#1{\textsf{#1}}\fi
\ifx \doiurl  \undefined \def \doiurl#1{\textsf{#1}}\fi
\ifx \betal  \undefined \def \betal{\textit{et al.}}\fi
\ifx \binstitute  \undefined \def \binstitute#1{#1}\fi
\ifx \binstitutionaled  \undefined \def \binstitutionaled#1{#1}\fi
\ifx \bctitle  \undefined \def \bctitle#1{#1}\fi
\ifx \beditor  \undefined \def \beditor#1{#1}\fi
\ifx \bpublisher  \undefined \def \bpublisher#1{#1}\fi
\ifx \bbtitle  \undefined \def \bbtitle#1{#1}\fi
\ifx \bedition  \undefined \def \bedition#1{#1}\fi
\ifx \bseriesno  \undefined \def \bseriesno#1{#1}\fi
\ifx \blocation  \undefined \def \blocation#1{#1}\fi
\ifx \bsertitle  \undefined \def \bsertitle#1{#1}\fi
\ifx \bsnm \undefined \def \bsnm#1{#1}\fi
\ifx \bsuffix \undefined \def \bsuffix#1{#1}\fi
\ifx \bparticle \undefined \def \bparticle#1{#1}\fi
\ifx \barticle \undefined \def \barticle#1{#1}\fi
\ifx \bconfdate \undefined \def \bconfdate #1{#1}\fi
\ifx \botherref \undefined \def \botherref #1{#1}\fi
\ifx \url \undefined \def \url#1{\textsf{#1}}\fi
\ifx \bchapter \undefined \def \bchapter#1{#1}\fi
\ifx \bbook \undefined \def \bbook#1{#1}\fi
\ifx \bcomment \undefined \def \bcomment#1{#1}\fi
\ifx \oauthor \undefined \def \oauthor#1{#1}\fi
\ifx \citeauthoryear \undefined \def \citeauthoryear#1{#1}\fi
\ifx \endbibitem  \undefined \def \endbibitem {}\fi
\ifx \bconflocation  \undefined \def \bconflocation#1{#1}\fi
\ifx \arxivurl  \undefined \def \arxivurl#1{\textsf{#1}}\fi
\csname PreBibitemsHook\endcsname

\bibitem{pastor2015epidemic}
\begin{barticle}
\bauthor{\bsnm{Pastor-Satorras}, \binits{R.}},
\bauthor{\bsnm{Castellano}, \binits{C.}},
\bauthor{\bsnm{Van~Mieghem}, \binits{P.}},
\bauthor{\bsnm{Vespignani}, \binits{A.}}:
\batitle{Epidemic processes in complex networks}.
\bjtitle{Reviews of modern physics}
\bvolume{87}(\bissue{3}),
\bfpage{925}
(\byear{2015})
\end{barticle}
\endbibitem

\bibitem{vosoughi2018spread}
\begin{barticle}
\bauthor{\bsnm{Vosoughi}, \binits{S.}},
\bauthor{\bsnm{Roy}, \binits{D.}},
\bauthor{\bsnm{Aral}, \binits{S.}}:
\batitle{The spread of true and false news online}.
\bjtitle{Science}
\bvolume{359}(\bissue{6380}),
\bfpage{1146}--\blpage{1151}
(\byear{2018})
\end{barticle}
\endbibitem

\bibitem{santolini2018predicting}
\begin{barticle}
\bauthor{\bsnm{Santolini}, \binits{M.}},
\bauthor{\bsnm{Barab{\'a}si}, \binits{A.-L.}}:
\batitle{Predicting perturbation patterns from the topology of biological
  networks}.
\bjtitle{Proceedings of the National Academy of Sciences}
\bvolume{115}(\bissue{27}),
\bfpage{6375}--\blpage{6383}
(\byear{2018})
\end{barticle}
\endbibitem

\bibitem{cohen2003efficient}
\begin{barticle}
\bauthor{\bsnm{Cohen}, \binits{R.}},
\bauthor{\bsnm{Havlin}, \binits{S.}},
\bauthor{\bsnm{Ben-Avraham}, \binits{D.}}:
\batitle{Efficient immunization strategies for computer networks and
  populations}.
\bjtitle{Physical review letters}
\bvolume{91}(\bissue{24}),
\bfpage{247901}
(\byear{2003})
\end{barticle}
\endbibitem

\bibitem{preciado2014optimal}
\begin{barticle}
\bauthor{\bsnm{Preciado}, \binits{V.M.}},
\bauthor{\bsnm{Zargham}, \binits{M.}},
\bauthor{\bsnm{Enyioha}, \binits{C.}},
\bauthor{\bsnm{Jadbabaie}, \binits{A.}},
\bauthor{\bsnm{Pappas}, \binits{G.J.}}:
\batitle{Optimal resource allocation for network protection against spreading
  processes}.
\bjtitle{IEEE Transactions on Control of Network Systems}
\bvolume{1}(\bissue{1}),
\bfpage{99}--\blpage{108}
(\byear{2014})
\end{barticle}
\endbibitem

\bibitem{wasserman1994social}
\begin{bbook}
\bauthor{\bsnm{Wasserman}, \binits{S.}},
\bauthor{\bsnm{Faust}, \binits{K.}}, \betal:
\bbtitle{Social Network Analysis: Methods and Applications}
vol. \bseriesno{8}.
\bpublisher{Cambridge university press}, \blocation{???}
(\byear{1994})
\end{bbook}
\endbibitem

\bibitem{freeman1977set}
\begin{botherref}
\oauthor{\bsnm{Freeman}, \binits{L.C.}}:
A set of measures of centrality based on betweenness.
Sociometry,
35--41
(1977)
\end{botherref}
\endbibitem

\bibitem{bonacich1972factoring}
\begin{barticle}
\bauthor{\bsnm{Bonacich}, \binits{P.}}:
\batitle{Factoring and weighting approaches to status scores and clique
  identification}.
\bjtitle{Journal of mathematical sociology}
\bvolume{2}(\bissue{1}),
\bfpage{113}--\blpage{120}
(\byear{1972})
\end{barticle}
\endbibitem

\bibitem{radicchi2016leveraging}
\begin{barticle}
\bauthor{\bsnm{Radicchi}, \binits{F.}},
\bauthor{\bsnm{Castellano}, \binits{C.}}:
\batitle{Leveraging percolation theory to single out influential spreaders in
  networks}.
\bjtitle{Physical Review E}
\bvolume{93}(\bissue{6}),
\bfpage{062314}
(\byear{2016})
\end{barticle}
\endbibitem

\bibitem{erkol2018influence}
\begin{barticle}
\bauthor{\bsnm{Erkol}, \binits{{\c{S}}.}},
\bauthor{\bsnm{Faqeeh}, \binits{A.}},
\bauthor{\bsnm{Radicchi}, \binits{F.}}:
\batitle{Influence maximization in noisy networks}.
\bjtitle{EPL (Europhysics Letters)}
\bvolume{123}(\bissue{5}),
\bfpage{58007}
(\byear{2018})
\end{barticle}
\endbibitem

\bibitem{groendyke2011bayesian}
\begin{barticle}
\bauthor{\bsnm{Groendyke}, \binits{C.}},
\bauthor{\bsnm{Welch}, \binits{D.}},
\bauthor{\bsnm{Hunter}, \binits{D.R.}}:
\batitle{Bayesian inference for contact networks given epidemic data}.
\bjtitle{Scandinavian Journal of Statistics}
\bvolume{38}(\bissue{3}),
\bfpage{600}--\blpage{616}
(\byear{2011})
\end{barticle}
\endbibitem

\bibitem{chinazzi2020effect}
\begin{barticle}
\bauthor{\bsnm{Chinazzi}, \binits{M.}},
\bauthor{\bsnm{Davis}, \binits{J.T.}},
\bauthor{\bsnm{Ajelli}, \binits{M.}},
\bauthor{\bsnm{Gioannini}, \binits{C.}},
\bauthor{\bsnm{Litvinova}, \binits{M.}},
\bauthor{\bsnm{Merler}, \binits{S.}},
\bauthor{\bparticle{y} \bsnm{Piontti}, \binits{A.P.}},
\bauthor{\bsnm{Mu}, \binits{K.}},
\bauthor{\bsnm{Rossi}, \binits{L.}},
\bauthor{\bsnm{Sun}, \binits{K.}}, \betal:
\batitle{The effect of travel restrictions on the spread of the 2019 novel
  coronavirus (covid-19) outbreak}.
\bjtitle{Science}
\bvolume{368}(\bissue{6489}),
\bfpage{395}--\blpage{400}
(\byear{2020})
\end{barticle}
\endbibitem

\bibitem{stack2013inferring}
\begin{barticle}
\bauthor{\bsnm{Stack}, \binits{J.C.}},
\bauthor{\bsnm{Bansal}, \binits{S.}},
\bauthor{\bsnm{Kumar}, \binits{V.A.}},
\bauthor{\bsnm{Grenfell}, \binits{B.}}:
\batitle{Inferring population-level contact heterogeneity from common epidemic
  data}.
\bjtitle{Journal of the Royal Society Interface}
\bvolume{10}(\bissue{78}),
\bfpage{20120578}
(\byear{2013})
\end{barticle}
\endbibitem

\bibitem{mishra2016impact}
\begin{barticle}
\bauthor{\bsnm{Mishra}, \binits{B.K.}},
\bauthor{\bsnm{Haldar}, \binits{K.}},
\bauthor{\bsnm{Sinha}, \binits{D.N.}}:
\batitle{Impact of information based classification on network epidemics}.
\bjtitle{Scientific reports}
\bvolume{6}(\bissue{1}),
\bfpage{1}--\blpage{17}
(\byear{2016})
\end{barticle}
\endbibitem

\bibitem{davis2020phase}
\begin{barticle}
\bauthor{\bsnm{Davis}, \binits{J.T.}},
\bauthor{\bsnm{Perra}, \binits{N.}},
\bauthor{\bsnm{Zhang}, \binits{Q.}},
\bauthor{\bsnm{Moreno}, \binits{Y.}},
\bauthor{\bsnm{Vespignani}, \binits{A.}}:
\batitle{Phase transitions in information spreading on structured populations}.
\bjtitle{Nature Physics}
\bvolume{16}(\bissue{5}),
\bfpage{590}--\blpage{596}
(\byear{2020})
\end{barticle}
\endbibitem

\bibitem{gomez2012inferring}
\begin{barticle}
\bauthor{\bsnm{Gomez-Rodriguez}, \binits{M.}},
\bauthor{\bsnm{Leskovec}, \binits{J.}},
\bauthor{\bsnm{Krause}, \binits{A.}}:
\batitle{Inferring networks of diffusion and influence}.
\bjtitle{ACM Transactions on Knowledge Discovery from Data (TKDD)}
\bvolume{5}(\bissue{4}),
\bfpage{1}--\blpage{37}
(\byear{2012})
\end{barticle}
\endbibitem

\bibitem{ellinas2016project}
\begin{barticle}
\bauthor{\bsnm{Ellinas}, \binits{C.}},
\bauthor{\bsnm{Allan}, \binits{N.}},
\bauthor{\bsnm{Johansson}, \binits{A.}}:
\batitle{Project systemic risk: Application examples of a network model}.
\bjtitle{International Journal of Production Economics}
\bvolume{182},
\bfpage{50}--\blpage{62}
(\byear{2016})
\end{barticle}
\endbibitem

\bibitem{vanhoucke2013overview}
\begin{botherref}
\oauthor{\bsnm{Vanhoucke}, \binits{M.}}:
An overview of recent research results and future research avenues using
  simulation studies in project management.
International Scholarly Research Notices
\textbf{2013}
(2013)
\end{botherref}
\endbibitem

\bibitem{santolini2020uncovering}
\begin{botherref}
\oauthor{\bsnm{Santolini}, \binits{M.}},
\oauthor{\bsnm{Ellinas}, \binits{C.}},
\oauthor{\bsnm{Nicolaides}, \binits{C.}}:
Uncovering the fragility of large-scale engineering projects
\end{botherref}
\endbibitem

\bibitem{valls2001criticality}
\begin{barticle}
\bauthor{\bsnm{Valls}, \binits{V.}},
\bauthor{\bsnm{Lino}, \binits{P.}}:
\batitle{Criticality analysis in activity-on-node networks with minimal time
  lags}.
\bjtitle{Annals of operations research}
\bvolume{102}(\bissue{1-4}),
\bfpage{17}--\blpage{37}
(\byear{2001})
\end{barticle}
\endbibitem

\bibitem{ellinas2015robust}
\begin{barticle}
\bauthor{\bsnm{Ellinas}, \binits{C.}},
\bauthor{\bsnm{Allan}, \binits{N.}},
\bauthor{\bsnm{Durugbo}, \binits{C.}},
\bauthor{\bsnm{Johansson}, \binits{A.}}:
\batitle{How robust is your project? from local failures to global
  catastrophes: A complex networks approach to project systemic risk}.
\bjtitle{PloS one}
\bvolume{10}(\bissue{11}),
\bfpage{0142469}
(\byear{2015})
\end{barticle}
\endbibitem

\bibitem{guo2019modeling}
\begin{barticle}
\bauthor{\bsnm{Guo}, \binits{N.}},
\bauthor{\bsnm{Guo}, \binits{P.}},
\bauthor{\bsnm{Dong}, \binits{H.}},
\bauthor{\bsnm{Zhao}, \binits{J.}},
\bauthor{\bsnm{Han}, \binits{Q.}}:
\batitle{Modeling and analysis of cascading failures in projects: A complex
  network approach}.
\bjtitle{Computers \& Industrial Engineering}
\bvolume{127},
\bfpage{1}--\blpage{7}
(\byear{2019})
\end{barticle}
\endbibitem

\bibitem{evrard2004boosting}
\begin{botherref}
\oauthor{\bsnm{Evrard}, \binits{D.}},
\oauthor{\bsnm{Nieto-Rodriguez}, \binits{A.}}:
Boosting business performance through programme and project management.
PriceWaterhouseCoopers
(2004)
\end{botherref}
\endbibitem

\bibitem{budzier2011your}
\begin{barticle}
\bauthor{\bsnm{Budzier}, \binits{A.}}, \betal:
\batitle{Why your it project may be riskier than you think}.
\bjtitle{Harvard Business Review}
\bvolume{89}(\bissue{9}),
\bfpage{23}--\blpage{25}
(\byear{2011})
\end{barticle}
\endbibitem

\bibitem{flyvbjerg2003common}
\begin{barticle}
\bauthor{\bsnm{Flyvbjerg}, \binits{B.}},
\bauthor{\bsnm{Skamris~Holm}, \binits{M.K.}},
\bauthor{\bsnm{Buhl}, \binits{S.L.}}:
\batitle{How common and how large are cost overruns in transport infrastructure
  projects?}
\bjtitle{Transport reviews}
\bvolume{23}(\bissue{1}),
\bfpage{71}--\blpage{88}
(\byear{2003})
\end{barticle}
\endbibitem

\bibitem{flyvbjerg2007cost}
\begin{barticle}
\bauthor{\bsnm{Flyvbjerg}, \binits{B.}}:
\batitle{Cost overruns and demand shortfalls in urban rail and other
  infrastructure}.
\bjtitle{Transportation Planning and Technology}
\bvolume{30}(\bissue{1}),
\bfpage{9}--\blpage{30}
(\byear{2007})
\end{barticle}
\endbibitem

\bibitem{sosa2014realizing}
\begin{barticle}
\bauthor{\bsnm{Sosa}, \binits{M.E.}}:
\batitle{Realizing the need for rework: From task interdependence to social
  networks}.
\bjtitle{Production and Operations Management}
\bvolume{23}(\bissue{8}),
\bfpage{1312}--\blpage{1331}
(\byear{2014})
\end{barticle}
\endbibitem

\bibitem{terwiesch1999managing}
\begin{barticle}
\bauthor{\bsnm{Terwiesch}, \binits{C.}},
\bauthor{\bsnm{Loch}, \binits{C.H.}}:
\batitle{Managing the process of engineering change orders: the case of the
  climate control system in automobile development}.
\bjtitle{Journal of Product Innovation Management: AN INTERNATIONAL PUBLICATION
  OF THE PRODUCT DEVELOPMENT \& MANAGEMENT ASSOCIATION}
\bvolume{16}(\bissue{2}),
\bfpage{160}--\blpage{172}
(\byear{1999})
\end{barticle}
\endbibitem

\bibitem{ellinas2019domino}
\begin{barticle}
\bauthor{\bsnm{Ellinas}, \binits{C.}}:
\batitle{The domino effect: An empirical exposition of systemic risk across
  project networks}.
\bjtitle{Production and Operations Management}
\bvolume{28}(\bissue{1}),
\bfpage{63}--\blpage{81}
(\byear{2019})
\end{barticle}
\endbibitem

\bibitem{mihm2003problem}
\begin{barticle}
\bauthor{\bsnm{Mihm}, \binits{J.}},
\bauthor{\bsnm{Loch}, \binits{C.}},
\bauthor{\bsnm{Huchzermeier}, \binits{A.}}:
\batitle{Problem--solving oscillations in complex engineering projects}.
\bjtitle{Management Science}
\bvolume{49}(\bissue{6}),
\bfpage{733}--\blpage{750}
(\byear{2003})
\end{barticle}
\endbibitem

\bibitem{wang2018development}
\begin{barticle}
\bauthor{\bsnm{Wang}, \binits{J.}},
\bauthor{\bsnm{Yang}, \binits{N.}},
\bauthor{\bsnm{Zhang}, \binits{Y.}},
\bauthor{\bsnm{Song}, \binits{Y.}}:
\batitle{Development of the mitigation strategy against the schedule risks of
  the r\&d project through controlling the cascading failure of the r\&d
  network}.
\bjtitle{Physica A: Statistical Mechanics and its Applications}
\bvolume{508},
\bfpage{390}--\blpage{401}
(\byear{2018})
\end{barticle}
\endbibitem

\bibitem{ellinas2018modelling}
\begin{barticle}
\bauthor{\bsnm{Ellinas}, \binits{C.}}:
\batitle{Modelling indirect interactions during failure spreading in a project
  activity network}.
\bjtitle{Scientific reports}
\bvolume{8}(\bissue{1}),
\bfpage{1}--\blpage{12}
(\byear{2018})
\end{barticle}
\endbibitem

\bibitem{estrada2010quantifying}
\begin{barticle}
\bauthor{\bsnm{Estrada}, \binits{E.}}:
\batitle{Quantifying network heterogeneity}.
\bjtitle{Physical Review E}
\bvolume{82}(\bissue{6}),
\bfpage{066102}
(\byear{2010})
\end{barticle}
\endbibitem

\bibitem{ye2013entropy}
\begin{bchapter}
\bauthor{\bsnm{Ye}, \binits{C.}},
\bauthor{\bsnm{Wilson}, \binits{R.C.}},
\bauthor{\bsnm{Comin}, \binits{C.H.}},
\bauthor{\bsnm{Costa}, \binits{L.d.F.}},
\bauthor{\bsnm{Hancock}, \binits{E.R.}}:
\bctitle{Entropy and heterogeneity measures for directed graphs}.
In: \bbtitle{International Workshop on Similarity-Based Pattern Recognition},
pp. \bfpage{219}--\blpage{234}
(\byear{2013}).
\bcomment{Springer}
\end{bchapter}
\endbibitem

\bibitem{moreno2002epidemic}
\begin{barticle}
\bauthor{\bsnm{Moreno}, \binits{Y.}},
\bauthor{\bsnm{Pastor-Satorras}, \binits{R.}},
\bauthor{\bsnm{Vespignani}, \binits{A.}}:
\batitle{Epidemic outbreaks in complex heterogeneous networks}.
\bjtitle{The European Physical Journal B-Condensed Matter and Complex Systems}
\bvolume{26}(\bissue{4}),
\bfpage{521}--\blpage{529}
(\byear{2002})
\end{barticle}
\endbibitem

\bibitem{xiao2018correlation}
\begin{barticle}
\bauthor{\bsnm{Xiao}, \binits{X.-m.}},
\bauthor{\bsnm{Jia}, \binits{L.-m.}},
\bauthor{\bsnm{Wang}, \binits{Y.-h.}}:
\batitle{Correlation between heterogeneity and vulnerability of subway networks
  based on passenger flow}.
\bjtitle{Journal of Rail Transport Planning \& Management}
\bvolume{8}(\bissue{2}),
\bfpage{145}--\blpage{157}
(\byear{2018})
\end{barticle}
\endbibitem

\bibitem{sun2016impact}
\begin{barticle}
\bauthor{\bsnm{Sun}, \binits{S.}},
\bauthor{\bsnm{Wu}, \binits{Y.}},
\bauthor{\bsnm{Ma}, \binits{Y.}},
\bauthor{\bsnm{Wang}, \binits{L.}},
\bauthor{\bsnm{Gao}, \binits{Z.}},
\bauthor{\bsnm{Xia}, \binits{C.}}:
\batitle{Impact of degree heterogeneity on attack vulnerability of
  interdependent networks}.
\bjtitle{Scientific reports}
\bvolume{6},
\bfpage{32983}
(\byear{2016})
\end{barticle}
\endbibitem

\bibitem{karsai2014time}
\begin{barticle}
\bauthor{\bsnm{Karsai}, \binits{M.}},
\bauthor{\bsnm{Perra}, \binits{N.}},
\bauthor{\bsnm{Vespignani}, \binits{A.}}:
\batitle{Time varying networks and the weakness of strong ties}.
\bjtitle{Scientific reports}
\bvolume{4},
\bfpage{4001}
(\byear{2014})
\end{barticle}
\endbibitem

\bibitem{sun2015contrasting}
\begin{barticle}
\bauthor{\bsnm{Sun}, \binits{K.}},
\bauthor{\bsnm{Baronchelli}, \binits{A.}},
\bauthor{\bsnm{Perra}, \binits{N.}}:
\batitle{Contrasting effects of strong ties on sir and sis processes in
  temporal networks}.
\bjtitle{The European Physical Journal B}
\bvolume{88}(\bissue{12}),
\bfpage{1}--\blpage{8}
(\byear{2015})
\end{barticle}
\endbibitem

\bibitem{perra2012activity}
\begin{barticle}
\bauthor{\bsnm{Perra}, \binits{N.}},
\bauthor{\bsnm{Gon{\c{c}}alves}, \binits{B.}},
\bauthor{\bsnm{Pastor-Satorras}, \binits{R.}},
\bauthor{\bsnm{Vespignani}, \binits{A.}}:
\batitle{Activity driven modeling of time varying networks}.
\bjtitle{Scientific reports}
\bvolume{2},
\bfpage{469}
(\byear{2012})
\end{barticle}
\endbibitem

\bibitem{liu2014controlling}
\begin{barticle}
\bauthor{\bsnm{Liu}, \binits{S.}},
\bauthor{\bsnm{Perra}, \binits{N.}},
\bauthor{\bsnm{Karsai}, \binits{M.}},
\bauthor{\bsnm{Vespignani}, \binits{A.}}:
\batitle{Controlling contagion processes in activity driven networks}.
\bjtitle{Phys. Rev. Lett.}
\bvolume{112},
\bfpage{118702}
(\byear{2014}).
doi:\doiurl{10.1103/PhysRevLett.112.118702}
\end{barticle}
\endbibitem

\bibitem{pozzana2017epidemic}
\begin{barticle}
\bauthor{\bsnm{Pozzana}, \binits{I.}},
\bauthor{\bsnm{Sun}, \binits{K.}},
\bauthor{\bsnm{Perra}, \binits{N.}}:
\batitle{Epidemic spreading on activity-driven networks with attractiveness}.
\bjtitle{Physical Review E}
\bvolume{96}(\bissue{4}),
\bfpage{042310}
(\byear{2017})
\end{barticle}
\endbibitem

\bibitem{ubaldi2017burstiness}
\begin{barticle}
\bauthor{\bsnm{Ubaldi}, \binits{E.}},
\bauthor{\bsnm{Vezzani}, \binits{A.}},
\bauthor{\bsnm{Karsai}, \binits{M.}},
\bauthor{\bsnm{Perra}, \binits{N.}},
\bauthor{\bsnm{Burioni}, \binits{R.}}:
\batitle{Burstiness and tie activation strategies in time-varying social
  networks}.
\bjtitle{Scientific Reports}
\bvolume{7},
\bfpage{46225}
(\byear{2017})
\end{barticle}
\endbibitem

\bibitem{nadini2018epidemic}
\begin{barticle}
\bauthor{\bsnm{Nadini}, \binits{M.}},
\bauthor{\bsnm{Sun}, \binits{K.}},
\bauthor{\bsnm{Ubaldi}, \binits{E.}},
\bauthor{\bsnm{Starnini}, \binits{M.}},
\bauthor{\bsnm{Rizzo}, \binits{A.}},
\bauthor{\bsnm{Perra}, \binits{N.}}:
\batitle{Epidemic spreading in modular time-varying networks}.
\bjtitle{Scientific reports}
\bvolume{8}(\bissue{1}),
\bfpage{1}--\blpage{11}
(\byear{2018})
\end{barticle}
\endbibitem

\bibitem{baccarini1996concept}
\begin{barticle}
\bauthor{\bsnm{Baccarini}, \binits{D.}}:
\batitle{The concept of project complexity—a review}.
\bjtitle{International journal of project management}
\bvolume{14}(\bissue{4}),
\bfpage{201}--\blpage{204}
(\byear{1996})
\end{barticle}
\endbibitem

\bibitem{jacobs2011product}
\begin{barticle}
\bauthor{\bsnm{Jacobs}, \binits{M.A.}},
\bauthor{\bsnm{Swink}, \binits{M.}}:
\batitle{Product portfolio architectural complexity and operational
  performance: Incorporating the roles of learning and fixed assets}.
\bjtitle{Journal of Operations Management}
\bvolume{29}(\bissue{7-8}),
\bfpage{677}--\blpage{691}
(\byear{2011})
\end{barticle}
\endbibitem

\bibitem{ellinas2016toward}
\begin{barticle}
\bauthor{\bsnm{Ellinas}, \binits{C.}},
\bauthor{\bsnm{Allan}, \binits{N.}},
\bauthor{\bsnm{Johansson}, \binits{A.}}:
\batitle{Toward project complexity evaluation: A structural perspective}.
\bjtitle{IEEE Systems Journal}
\bvolume{12}(\bissue{1}),
\bfpage{228}--\blpage{239}
(\byear{2016})
\end{barticle}
\endbibitem

\bibitem{lawyer2015understanding}
\begin{barticle}
\bauthor{\bsnm{Lawyer}, \binits{G.}}:
\batitle{Understanding the influence of all nodes in a network}.
\bjtitle{Scientific reports}
\bvolume{5}(\bissue{1}),
\bfpage{1}--\blpage{9}
(\byear{2015})
\end{barticle}
\endbibitem

\end{thebibliography}

\newcommand{\BMCxmlcomment}[1]{}

\BMCxmlcomment{

<refgrp>

<bibl id="B1">
  <title><p>Epidemic processes in complex networks</p></title>
  <aug>
    <au><snm>Pastor Satorras</snm><fnm>R</fnm></au>
    <au><snm>Castellano</snm><fnm>C</fnm></au>
    <au><snm>Van Mieghem</snm><fnm>P</fnm></au>
    <au><snm>Vespignani</snm><fnm>A</fnm></au>
  </aug>
  <source>Reviews of modern physics</source>
  <publisher>APS</publisher>
  <pubdate>2015</pubdate>
  <volume>87</volume>
  <issue>3</issue>
  <fpage>925</fpage>
</bibl>

<bibl id="B2">
  <title><p>The spread of true and false news online</p></title>
  <aug>
    <au><snm>Vosoughi</snm><fnm>S</fnm></au>
    <au><snm>Roy</snm><fnm>D</fnm></au>
    <au><snm>Aral</snm><fnm>S</fnm></au>
  </aug>
  <source>Science</source>
  <publisher>American Association for the Advancement of Science</publisher>
  <pubdate>2018</pubdate>
  <volume>359</volume>
  <issue>6380</issue>
  <fpage>1146</fpage>
  <lpage>-1151</lpage>
</bibl>

<bibl id="B3">
  <title><p>Predicting perturbation patterns from the topology of biological
  networks</p></title>
  <aug>
    <au><snm>Santolini</snm><fnm>M</fnm></au>
    <au><snm>Barab{\'a}si</snm><fnm>AL</fnm></au>
  </aug>
  <source>Proceedings of the National Academy of Sciences</source>
  <publisher>National Acad Sciences</publisher>
  <pubdate>2018</pubdate>
  <volume>115</volume>
  <issue>27</issue>
  <fpage>E6375</fpage>
  <lpage>-E6383</lpage>
</bibl>

<bibl id="B4">
  <title><p>Efficient immunization strategies for computer networks and
  populations</p></title>
  <aug>
    <au><snm>Cohen</snm><fnm>R</fnm></au>
    <au><snm>Havlin</snm><fnm>S</fnm></au>
    <au><snm>Ben Avraham</snm><fnm>D</fnm></au>
  </aug>
  <source>Physical review letters</source>
  <publisher>APS</publisher>
  <pubdate>2003</pubdate>
  <volume>91</volume>
  <issue>24</issue>
  <fpage>247901</fpage>
</bibl>

<bibl id="B5">
  <title><p>Optimal resource allocation for network protection against
  spreading processes</p></title>
  <aug>
    <au><snm>Preciado</snm><fnm>VM</fnm></au>
    <au><snm>Zargham</snm><fnm>M</fnm></au>
    <au><snm>Enyioha</snm><fnm>C</fnm></au>
    <au><snm>Jadbabaie</snm><fnm>A</fnm></au>
    <au><snm>Pappas</snm><fnm>GJ</fnm></au>
  </aug>
  <source>IEEE Transactions on Control of Network Systems</source>
  <publisher>IEEE</publisher>
  <pubdate>2014</pubdate>
  <volume>1</volume>
  <issue>1</issue>
  <fpage>99</fpage>
  <lpage>-108</lpage>
</bibl>

<bibl id="B6">
  <title><p>Social network analysis: Methods and applications</p></title>
  <aug>
    <au><snm>Wasserman</snm><fnm>S</fnm></au>
    <au><snm>Faust</snm><fnm>K</fnm></au>
    <au><cnm>others</cnm></au>
  </aug>
  <publisher>Cambridge university press</publisher>
  <pubdate>1994</pubdate>
  <volume>8</volume>
</bibl>

<bibl id="B7">
  <title><p>A set of measures of centrality based on betweenness</p></title>
  <aug>
    <au><snm>Freeman</snm><fnm>LC</fnm></au>
  </aug>
  <source>Sociometry</source>
  <publisher>JSTOR</publisher>
  <pubdate>1977</pubdate>
  <fpage>35</fpage>
  <lpage>-41</lpage>
</bibl>

<bibl id="B8">
  <title><p>Factoring and weighting approaches to status scores and clique
  identification</p></title>
  <aug>
    <au><snm>Bonacich</snm><fnm>P</fnm></au>
  </aug>
  <source>Journal of mathematical sociology</source>
  <publisher>Taylor \& Francis</publisher>
  <pubdate>1972</pubdate>
  <volume>2</volume>
  <issue>1</issue>
  <fpage>113</fpage>
  <lpage>-120</lpage>
</bibl>

<bibl id="B9">
  <title><p>Leveraging percolation theory to single out influential spreaders
  in networks</p></title>
  <aug>
    <au><snm>Radicchi</snm><fnm>F</fnm></au>
    <au><snm>Castellano</snm><fnm>C</fnm></au>
  </aug>
  <source>Physical Review E</source>
  <publisher>APS</publisher>
  <pubdate>2016</pubdate>
  <volume>93</volume>
  <issue>6</issue>
  <fpage>062314</fpage>
</bibl>

<bibl id="B10">
  <title><p>Influence maximization in noisy networks</p></title>
  <aug>
    <au><snm>Erkol</snm><fnm>{\c{S}}</fnm></au>
    <au><snm>Faqeeh</snm><fnm>A</fnm></au>
    <au><snm>Radicchi</snm><fnm>F</fnm></au>
  </aug>
  <source>EPL (Europhysics Letters)</source>
  <publisher>IOP Publishing</publisher>
  <pubdate>2018</pubdate>
  <volume>123</volume>
  <issue>5</issue>
  <fpage>58007</fpage>
</bibl>

<bibl id="B11">
  <title><p>Bayesian inference for contact networks given epidemic
  data</p></title>
  <aug>
    <au><snm>Groendyke</snm><fnm>C</fnm></au>
    <au><snm>Welch</snm><fnm>D</fnm></au>
    <au><snm>Hunter</snm><fnm>DR</fnm></au>
  </aug>
  <source>Scandinavian Journal of Statistics</source>
  <publisher>Wiley Online Library</publisher>
  <pubdate>2011</pubdate>
  <volume>38</volume>
  <issue>3</issue>
  <fpage>600</fpage>
  <lpage>-616</lpage>
</bibl>

<bibl id="B12">
  <title><p>The effect of travel restrictions on the spread of the 2019 novel
  coronavirus (COVID-19) outbreak</p></title>
  <aug>
    <au><snm>Chinazzi</snm><fnm>M</fnm></au>
    <au><snm>Davis</snm><fnm>JT</fnm></au>
    <au><snm>Ajelli</snm><fnm>M</fnm></au>
    <au><snm>Gioannini</snm><fnm>C</fnm></au>
    <au><snm>Litvinova</snm><fnm>M</fnm></au>
    <au><snm>Merler</snm><fnm>S</fnm></au>
    <au><snm>Piontti</snm><fnm>AP</fnm></au>
    <au><snm>Mu</snm><fnm>K</fnm></au>
    <au><snm>Rossi</snm><fnm>L</fnm></au>
    <au><snm>Sun</snm><fnm>K</fnm></au>
    <au><cnm>others</cnm></au>
  </aug>
  <source>Science</source>
  <publisher>American Association for the Advancement of Science</publisher>
  <pubdate>2020</pubdate>
  <volume>368</volume>
  <issue>6489</issue>
  <fpage>395</fpage>
  <lpage>-400</lpage>
</bibl>

<bibl id="B13">
  <title><p>Inferring population-level contact heterogeneity from common
  epidemic data</p></title>
  <aug>
    <au><snm>Stack</snm><fnm>JC</fnm></au>
    <au><snm>Bansal</snm><fnm>S</fnm></au>
    <au><snm>Kumar</snm><fnm>VA</fnm></au>
    <au><snm>Grenfell</snm><fnm>B</fnm></au>
  </aug>
  <source>Journal of the Royal Society Interface</source>
  <publisher>The Royal Society</publisher>
  <pubdate>2013</pubdate>
  <volume>10</volume>
  <issue>78</issue>
  <fpage>20120578</fpage>
</bibl>

<bibl id="B14">
  <title><p>Impact of information based classification on network
  epidemics</p></title>
  <aug>
    <au><snm>Mishra</snm><fnm>BK</fnm></au>
    <au><snm>Haldar</snm><fnm>K</fnm></au>
    <au><snm>Sinha</snm><fnm>DN</fnm></au>
  </aug>
  <source>Scientific reports</source>
  <publisher>Nature Publishing Group</publisher>
  <pubdate>2016</pubdate>
  <volume>6</volume>
  <issue>1</issue>
  <fpage>1</fpage>
  <lpage>-17</lpage>
</bibl>

<bibl id="B15">
  <title><p>Phase transitions in information spreading on structured
  populations</p></title>
  <aug>
    <au><snm>Davis</snm><fnm>JT</fnm></au>
    <au><snm>Perra</snm><fnm>N</fnm></au>
    <au><snm>Zhang</snm><fnm>Q</fnm></au>
    <au><snm>Moreno</snm><fnm>Y</fnm></au>
    <au><snm>Vespignani</snm><fnm>A</fnm></au>
  </aug>
  <source>Nature Physics</source>
  <publisher>Nature Publishing Group</publisher>
  <pubdate>2020</pubdate>
  <volume>16</volume>
  <issue>5</issue>
  <fpage>590</fpage>
  <lpage>-596</lpage>
</bibl>

<bibl id="B16">
  <title><p>Inferring networks of diffusion and influence</p></title>
  <aug>
    <au><snm>Gomez Rodriguez</snm><fnm>M</fnm></au>
    <au><snm>Leskovec</snm><fnm>J</fnm></au>
    <au><snm>Krause</snm><fnm>A</fnm></au>
  </aug>
  <source>ACM Transactions on Knowledge Discovery from Data (TKDD)</source>
  <publisher>ACM New York, NY, USA</publisher>
  <pubdate>2012</pubdate>
  <volume>5</volume>
  <issue>4</issue>
  <fpage>1</fpage>
  <lpage>-37</lpage>
</bibl>

<bibl id="B17">
  <title><p>Project systemic risk: Application examples of a network
  model</p></title>
  <aug>
    <au><snm>Ellinas</snm><fnm>C</fnm></au>
    <au><snm>Allan</snm><fnm>N</fnm></au>
    <au><snm>Johansson</snm><fnm>A</fnm></au>
  </aug>
  <source>International Journal of Production Economics</source>
  <publisher>Elsevier</publisher>
  <pubdate>2016</pubdate>
  <volume>182</volume>
  <fpage>50</fpage>
  <lpage>-62</lpage>
</bibl>

<bibl id="B18">
  <title><p>An overview of recent research results and future research avenues
  using simulation studies in project management</p></title>
  <aug>
    <au><snm>Vanhoucke</snm><fnm>M</fnm></au>
  </aug>
  <source>International Scholarly Research Notices</source>
  <publisher>Hindawi</publisher>
  <pubdate>2013</pubdate>
  <volume>2013</volume>
</bibl>

<bibl id="B19">
  <title><p>Uncovering the fragility of large-scale engineering
  projects</p></title>
  <aug>
    <au><snm>Santolini</snm><fnm>M</fnm></au>
    <au><snm>Ellinas</snm><fnm>C</fnm></au>
    <au><snm>Nicolaides</snm><fnm>C</fnm></au>
  </aug>
</bibl>

<bibl id="B20">
  <title><p>Criticality analysis in activity-on-node networks with minimal time
  lags</p></title>
  <aug>
    <au><snm>Valls</snm><fnm>V</fnm></au>
    <au><snm>Lino</snm><fnm>P</fnm></au>
  </aug>
  <source>Annals of operations research</source>
  <publisher>Springer</publisher>
  <pubdate>2001</pubdate>
  <volume>102</volume>
  <issue>1-4</issue>
  <fpage>17</fpage>
  <lpage>-37</lpage>
</bibl>

<bibl id="B21">
  <title><p>How robust is your project? From local failures to global
  catastrophes: A complex networks approach to project systemic
  risk</p></title>
  <aug>
    <au><snm>Ellinas</snm><fnm>C</fnm></au>
    <au><snm>Allan</snm><fnm>N</fnm></au>
    <au><snm>Durugbo</snm><fnm>C</fnm></au>
    <au><snm>Johansson</snm><fnm>A</fnm></au>
  </aug>
  <source>PloS one</source>
  <publisher>Public Library of Science San Francisco, CA USA</publisher>
  <pubdate>2015</pubdate>
  <volume>10</volume>
  <issue>11</issue>
  <fpage>e0142469</fpage>
</bibl>

<bibl id="B22">
  <title><p>Modeling and analysis of cascading failures in projects: A complex
  network approach</p></title>
  <aug>
    <au><snm>Guo</snm><fnm>N</fnm></au>
    <au><snm>Guo</snm><fnm>P</fnm></au>
    <au><snm>Dong</snm><fnm>H</fnm></au>
    <au><snm>Zhao</snm><fnm>J</fnm></au>
    <au><snm>Han</snm><fnm>Q</fnm></au>
  </aug>
  <source>Computers \& Industrial Engineering</source>
  <publisher>Elsevier</publisher>
  <pubdate>2019</pubdate>
  <volume>127</volume>
  <fpage>1</fpage>
  <lpage>-7</lpage>
</bibl>

<bibl id="B23">
  <title><p>Boosting business performance through programme and project
  management</p></title>
  <aug>
    <au><snm>Evrard</snm><fnm>D</fnm></au>
    <au><snm>Nieto Rodriguez</snm><fnm>A</fnm></au>
  </aug>
  <publisher>PriceWaterhouseCoopers</publisher>
  <pubdate>2004</pubdate>
</bibl>

<bibl id="B24">
  <title><p>Why your IT project may be riskier than you think</p></title>
  <aug>
    <au><snm>Budzier</snm><fnm>A</fnm></au>
    <au><cnm>others</cnm></au>
  </aug>
  <source>Harvard Business Review</source>
  <pubdate>2011</pubdate>
  <volume>89</volume>
  <issue>9</issue>
  <fpage>23</fpage>
  <lpage>-25</lpage>
</bibl>

<bibl id="B25">
  <title><p>How common and how large are cost overruns in transport
  infrastructure projects?</p></title>
  <aug>
    <au><snm>Flyvbjerg</snm><fnm>B</fnm></au>
    <au><snm>Skamris Holm</snm><fnm>MK</fnm></au>
    <au><snm>Buhl</snm><fnm>SL</fnm></au>
  </aug>
  <source>Transport reviews</source>
  <publisher>Taylor \& Francis</publisher>
  <pubdate>2003</pubdate>
  <volume>23</volume>
  <issue>1</issue>
  <fpage>71</fpage>
  <lpage>-88</lpage>
</bibl>

<bibl id="B26">
  <title><p>Cost overruns and demand shortfalls in urban rail and other
  infrastructure</p></title>
  <aug>
    <au><snm>Flyvbjerg</snm><fnm>B</fnm></au>
  </aug>
  <source>Transportation Planning and Technology</source>
  <publisher>Taylor \& Francis</publisher>
  <pubdate>2007</pubdate>
  <volume>30</volume>
  <issue>1</issue>
  <fpage>9</fpage>
  <lpage>-30</lpage>
</bibl>

<bibl id="B27">
  <title><p>Realizing the need for rework: From task interdependence to social
  networks</p></title>
  <aug>
    <au><snm>Sosa</snm><fnm>ME</fnm></au>
  </aug>
  <source>Production and Operations Management</source>
  <publisher>Wiley Online Library</publisher>
  <pubdate>2014</pubdate>
  <volume>23</volume>
  <issue>8</issue>
  <fpage>1312</fpage>
  <lpage>-1331</lpage>
</bibl>

<bibl id="B28">
  <title><p>Managing the process of engineering change orders: the case of the
  climate control system in automobile development</p></title>
  <aug>
    <au><snm>Terwiesch</snm><fnm>C</fnm></au>
    <au><snm>Loch</snm><fnm>CH</fnm></au>
  </aug>
  <source>Journal of Product Innovation Management: AN INTERNATIONAL
  PUBLICATION OF THE PRODUCT DEVELOPMENT \& MANAGEMENT ASSOCIATION</source>
  <publisher>Wiley Online Library</publisher>
  <pubdate>1999</pubdate>
  <volume>16</volume>
  <issue>2</issue>
  <fpage>160</fpage>
  <lpage>-172</lpage>
</bibl>

<bibl id="B29">
  <title><p>The domino effect: An empirical exposition of systemic risk across
  project networks</p></title>
  <aug>
    <au><snm>Ellinas</snm><fnm>C</fnm></au>
  </aug>
  <source>Production and Operations Management</source>
  <publisher>Wiley Online Library</publisher>
  <pubdate>2019</pubdate>
  <volume>28</volume>
  <issue>1</issue>
  <fpage>63</fpage>
  <lpage>-81</lpage>
</bibl>

<bibl id="B30">
  <title><p>Problem--solving oscillations in complex engineering
  projects</p></title>
  <aug>
    <au><snm>Mihm</snm><fnm>J</fnm></au>
    <au><snm>Loch</snm><fnm>C</fnm></au>
    <au><snm>Huchzermeier</snm><fnm>A</fnm></au>
  </aug>
  <source>Management Science</source>
  <publisher>INFORMS</publisher>
  <pubdate>2003</pubdate>
  <volume>49</volume>
  <issue>6</issue>
  <fpage>733</fpage>
  <lpage>-750</lpage>
</bibl>

<bibl id="B31">
  <title><p>Development of the mitigation strategy against the schedule risks
  of the R&ampD project through controlling the cascading failure of the R&ampD
  network</p></title>
  <aug>
    <au><snm>Wang</snm><fnm>J</fnm></au>
    <au><snm>Yang</snm><fnm>N</fnm></au>
    <au><snm>Zhang</snm><fnm>Y</fnm></au>
    <au><snm>Song</snm><fnm>Y</fnm></au>
  </aug>
  <source>Physica A: Statistical Mechanics and its Applications</source>
  <publisher>Elsevier</publisher>
  <pubdate>2018</pubdate>
  <volume>508</volume>
  <fpage>390</fpage>
  <lpage>-401</lpage>
</bibl>

<bibl id="B32">
  <title><p>Modelling indirect interactions during failure spreading in a
  project activity network</p></title>
  <aug>
    <au><snm>Ellinas</snm><fnm>C</fnm></au>
  </aug>
  <source>Scientific reports</source>
  <publisher>Nature Publishing Group</publisher>
  <pubdate>2018</pubdate>
  <volume>8</volume>
  <issue>1</issue>
  <fpage>1</fpage>
  <lpage>-12</lpage>
</bibl>

<bibl id="B33">
  <title><p>Quantifying network heterogeneity</p></title>
  <aug>
    <au><snm>Estrada</snm><fnm>E</fnm></au>
  </aug>
  <source>Physical Review E</source>
  <publisher>APS</publisher>
  <pubdate>2010</pubdate>
  <volume>82</volume>
  <issue>6</issue>
  <fpage>066102</fpage>
</bibl>

<bibl id="B34">
  <title><p>Entropy and heterogeneity measures for directed graphs</p></title>
  <aug>
    <au><snm>Ye</snm><fnm>C</fnm></au>
    <au><snm>Wilson</snm><fnm>RC</fnm></au>
    <au><snm>Comin</snm><fnm>CH</fnm></au>
    <au><snm>Costa</snm><fnm>LdF</fnm></au>
    <au><snm>Hancock</snm><fnm>ER</fnm></au>
  </aug>
  <source>International Workshop on Similarity-Based Pattern
  Recognition</source>
  <pubdate>2013</pubdate>
  <fpage>219</fpage>
  <lpage>-234</lpage>
</bibl>

<bibl id="B35">
  <title><p>Epidemic outbreaks in complex heterogeneous networks</p></title>
  <aug>
    <au><snm>Moreno</snm><fnm>Y</fnm></au>
    <au><snm>Pastor Satorras</snm><fnm>R</fnm></au>
    <au><snm>Vespignani</snm><fnm>A</fnm></au>
  </aug>
  <source>The European Physical Journal B-Condensed Matter and Complex
  Systems</source>
  <publisher>Springer</publisher>
  <pubdate>2002</pubdate>
  <volume>26</volume>
  <issue>4</issue>
  <fpage>521</fpage>
  <lpage>-529</lpage>
</bibl>

<bibl id="B36">
  <title><p>Correlation between heterogeneity and vulnerability of subway
  networks based on passenger flow</p></title>
  <aug>
    <au><snm>Xiao</snm><fnm>Xm</fnm></au>
    <au><snm>Jia</snm><fnm>Lm</fnm></au>
    <au><snm>Wang</snm><fnm>Yh</fnm></au>
  </aug>
  <source>Journal of Rail Transport Planning \& Management</source>
  <publisher>Elsevier</publisher>
  <pubdate>2018</pubdate>
  <volume>8</volume>
  <issue>2</issue>
  <fpage>145</fpage>
  <lpage>-157</lpage>
</bibl>

<bibl id="B37">
  <title><p>Impact of degree heterogeneity on attack vulnerability of
  interdependent networks</p></title>
  <aug>
    <au><snm>Sun</snm><fnm>S</fnm></au>
    <au><snm>Wu</snm><fnm>Y</fnm></au>
    <au><snm>Ma</snm><fnm>Y</fnm></au>
    <au><snm>Wang</snm><fnm>L</fnm></au>
    <au><snm>Gao</snm><fnm>Z</fnm></au>
    <au><snm>Xia</snm><fnm>C</fnm></au>
  </aug>
  <source>Scientific reports</source>
  <publisher>Nature Publishing Group</publisher>
  <pubdate>2016</pubdate>
  <volume>6</volume>
  <fpage>32983</fpage>
</bibl>

<bibl id="B38">
  <title><p>Time varying networks and the weakness of strong ties</p></title>
  <aug>
    <au><snm>Karsai</snm><fnm>M</fnm></au>
    <au><snm>Perra</snm><fnm>N</fnm></au>
    <au><snm>Vespignani</snm><fnm>A</fnm></au>
  </aug>
  <source>Scientific reports</source>
  <publisher>Nature Publishing Group</publisher>
  <pubdate>2014</pubdate>
  <volume>4</volume>
  <fpage>4001</fpage>
</bibl>

<bibl id="B39">
  <title><p>Contrasting effects of strong ties on SIR and SIS processes in
  temporal networks</p></title>
  <aug>
    <au><snm>Sun</snm><fnm>K</fnm></au>
    <au><snm>Baronchelli</snm><fnm>A</fnm></au>
    <au><snm>Perra</snm><fnm>N</fnm></au>
  </aug>
  <source>The European Physical Journal B</source>
  <publisher>Springer</publisher>
  <pubdate>2015</pubdate>
  <volume>88</volume>
  <issue>12</issue>
  <fpage>1</fpage>
  <lpage>-8</lpage>
</bibl>

<bibl id="B40">
  <title><p>Activity driven modeling of time varying networks</p></title>
  <aug>
    <au><snm>Perra</snm><fnm>N</fnm></au>
    <au><snm>Gon{\c{c}}alves</snm><fnm>B</fnm></au>
    <au><snm>Pastor Satorras</snm><fnm>R</fnm></au>
    <au><snm>Vespignani</snm><fnm>A</fnm></au>
  </aug>
  <source>Scientific reports</source>
  <publisher>Nature Publishing Group</publisher>
  <pubdate>2012</pubdate>
  <volume>2</volume>
  <fpage>469</fpage>
</bibl>

<bibl id="B41">
  <title><p>Controlling Contagion Processes in Activity Driven
  Networks</p></title>
  <aug>
    <au><snm>Liu</snm><fnm>S</fnm></au>
    <au><snm>Perra</snm><fnm>N</fnm></au>
    <au><snm>Karsai</snm><fnm>M</fnm></au>
    <au><snm>Vespignani</snm><fnm>A</fnm></au>
  </aug>
  <source>Phys. Rev. Lett.</source>
  <publisher>American Physical Society</publisher>
  <pubdate>2014</pubdate>
  <volume>112</volume>
  <fpage>118702</fpage>
  <url>https://link.aps.org/doi/10.1103/PhysRevLett.112.118702</url>
</bibl>

<bibl id="B42">
  <title><p>Epidemic spreading on activity-driven networks with
  attractiveness</p></title>
  <aug>
    <au><snm>Pozzana</snm><fnm>I</fnm></au>
    <au><snm>Sun</snm><fnm>K</fnm></au>
    <au><snm>Perra</snm><fnm>N</fnm></au>
  </aug>
  <source>Physical Review E</source>
  <publisher>APS</publisher>
  <pubdate>2017</pubdate>
  <volume>96</volume>
  <issue>4</issue>
  <fpage>042310</fpage>
</bibl>

<bibl id="B43">
  <title><p>Burstiness and tie activation strategies in time-varying social
  networks</p></title>
  <aug>
    <au><snm>Ubaldi</snm><fnm>E</fnm></au>
    <au><snm>Vezzani</snm><fnm>A</fnm></au>
    <au><snm>Karsai</snm><fnm>M</fnm></au>
    <au><snm>Perra</snm><fnm>N</fnm></au>
    <au><snm>Burioni</snm><fnm>R</fnm></au>
  </aug>
  <source>Scientific Reports</source>
  <publisher>Nature Publishing Group</publisher>
  <pubdate>2017</pubdate>
  <volume>7</volume>
  <fpage>46225</fpage>
</bibl>

<bibl id="B44">
  <title><p>Epidemic spreading in modular time-varying networks</p></title>
  <aug>
    <au><snm>Nadini</snm><fnm>M</fnm></au>
    <au><snm>Sun</snm><fnm>K</fnm></au>
    <au><snm>Ubaldi</snm><fnm>E</fnm></au>
    <au><snm>Starnini</snm><fnm>M</fnm></au>
    <au><snm>Rizzo</snm><fnm>A</fnm></au>
    <au><snm>Perra</snm><fnm>N</fnm></au>
  </aug>
  <source>Scientific reports</source>
  <publisher>Nature Publishing Group</publisher>
  <pubdate>2018</pubdate>
  <volume>8</volume>
  <issue>1</issue>
  <fpage>1</fpage>
  <lpage>-11</lpage>
</bibl>

<bibl id="B45">
  <title><p>The concept of project complexity—a review</p></title>
  <aug>
    <au><snm>Baccarini</snm><fnm>D</fnm></au>
  </aug>
  <source>International journal of project management</source>
  <publisher>Elsevier</publisher>
  <pubdate>1996</pubdate>
  <volume>14</volume>
  <issue>4</issue>
  <fpage>201</fpage>
  <lpage>-204</lpage>
</bibl>

<bibl id="B46">
  <title><p>Product portfolio architectural complexity and operational
  performance: Incorporating the roles of learning and fixed assets</p></title>
  <aug>
    <au><snm>Jacobs</snm><fnm>MA</fnm></au>
    <au><snm>Swink</snm><fnm>M</fnm></au>
  </aug>
  <source>Journal of Operations Management</source>
  <publisher>Elsevier</publisher>
  <pubdate>2011</pubdate>
  <volume>29</volume>
  <issue>7-8</issue>
  <fpage>677</fpage>
  <lpage>-691</lpage>
</bibl>

<bibl id="B47">
  <title><p>Toward project complexity evaluation: A structural
  perspective</p></title>
  <aug>
    <au><snm>Ellinas</snm><fnm>C</fnm></au>
    <au><snm>Allan</snm><fnm>N</fnm></au>
    <au><snm>Johansson</snm><fnm>A</fnm></au>
  </aug>
  <source>IEEE Systems Journal</source>
  <publisher>IEEE</publisher>
  <pubdate>2016</pubdate>
  <volume>12</volume>
  <issue>1</issue>
  <fpage>228</fpage>
  <lpage>-239</lpage>
</bibl>

<bibl id="B48">
  <title><p>Understanding the influence of all nodes in a network</p></title>
  <aug>
    <au><snm>Lawyer</snm><fnm>G</fnm></au>
  </aug>
  <source>Scientific reports</source>
  <publisher>Nature Publishing Group</publisher>
  <pubdate>2015</pubdate>
  <volume>5</volume>
  <issue>1</issue>
  <fpage>1</fpage>
  <lpage>-9</lpage>
</bibl>

</refgrp>
} 







\end{backmatter}
\end{document}